\documentclass[reprint,superscriptaddress, amsmath,amssymb,aps,longbibliography]{revtex4-1}

 \usepackage[bottom]{footmisc}

\usepackage[T1]{fontenc}
\usepackage[latin9]{inputenc}
\usepackage{mathrsfs}
\usepackage[maxfloats=256]{morefloats}
\usepackage{bm}
\usepackage{amsmath}
\usepackage{amsthm}
\usepackage{amssymb}
\usepackage{stmaryrd}
\usepackage{mathtools}
\usepackage{dsfont}

\edef\ordinarycolon{\mathchar\the\mathcode`: }
\edef\ordinaryequals{\mathchar\the\mathcode`= }

\usepackage{breqn}
\catcode`^=7

\AtBeginDocument{%
  \catcode`^=12
}

\makeatletter

\allowdisplaybreaks

\let\cat@comma@active\@empty

\usepackage[capitalize]{cleveref}

\usepackage{color}
\RequirePackage[dvipsnames,usenames]{xcolor}

\newif\ifnotes
\notestrue

\makeatother

\newcommand{\ba}{\begin{eqnarray}}
\newcommand{\ea}{\end{eqnarray}}

\begin{document}

\title{The Past as a Stochastic Process}

\author{David H. Wolpert}%
\affiliation{Santa Fe Institute}
\email{david.h.wolpert@gmail.com}
\homepage{ http://santafe.edu/~dhw}
\affiliation{Complexity Science Hub Vienna}
\affiliation{Arizona State University}
\author{Michael H. Price}
 \affiliation{Santa Fe Institute}
\author{Stefani A. Crabtree}
 \affiliation{Santa Fe Institute}
 \affiliation{Utah State University}
\author{Timothy A.\ Kohler}
 \affiliation{Santa Fe Institute}
 \affiliation{Washington State University}
\affiliation{Crow Canyon Archaeological Center}
\author{J\"urgen Jost}
 \affiliation{Santa Fe Institute}
 \affiliation{The Max Planck Institute for Mathematics in the Science}
\author{James Evans}
 \affiliation{Santa Fe Institute}
 \affiliation{University of Chicago}
\author{Peter F.\ Stadler}
 \affiliation{Santa Fe Institute}
 \affiliation{Leipzig University}
 \affiliation{The Max Planck Institute for Mathematics in the Science}
\author{Hajime Shimao}
 \affiliation{Santa Fe Institute}
\author{Manfred D.\ Laubichler}
 \affiliation{Santa Fe Institute}
\affiliation{Arizona State University}



\begin{abstract}
Historical processes manifest remarkable diversity. Nevertheless, scholars have long attempted to identify patterns and categorize historical actors and influences with some success. A stochastic process framework provides a structured approach for the analysis of large historical datasets that allows for detection of sometimes surprising patterns, identification of relevant causal actors both endogenous and exogenous to the process, and comparison between different historical cases. The combination of data, analytical tools and the organizing theoretical framework of stochastic processes complements traditional narrative approaches in history and archaeology.
\end{abstract}

\maketitle


%




How does history unfold? Are the dynamics of history determined by the agency of individual actors? Or, in contrast, are they determined by broad-scale processes that agents must react to but cannot affect? Is there some way to unify these opposites and account for a possible combination of factors? Social scientists have not found it easy to find a consensus on such fundamental issues. In the 19\textsuperscript{th} century, partly building on the work of Kant, thinkers like Hegel, Comte, Marx, Morgan, and Tylor emphasized necessity, generality, natural laws, and impersonal trends \cite{McAllister02}. They saw history as progressing in a law-like manner through well-defined stages, to approach some final destination. In the 20\textsuperscript{th} century, while still ascribing a prominent role to impersonal trends, authors like Spengler speculated about the inevitable decline of societies, removing the assumption of ``inherent progress'' made in those 19\textsuperscript{th}-century investigations. Both of these perspectives view impersonal processes --- what contemporary social scientists refer to as ``structural factors''  \cite{Sewell92} --- as key for understanding how history ultimately unfolds. In contrast, others like Dilthey and Simmel, and continuing through Geertz and  other late-20\textsuperscript{th}-century anthropologists, have argued that understanding human societies requires particular attention be given to individual agency, which, it is claimed, cannot be reduced to impersonal processes. 

Some historians have attempted to mediate this dichotomy by emphasizing interactions among individual actions and structural contexts -- linguistic, material, economic, etc. -- in detailed narratives. 
More recently environmental factors have been emphasized as well, both more gradual ones , such as climate change, which transform environmental variables (e.g., \cite{Weiss995}, \cite{Strawhacker20}) and unpredictable, exogenous events such as volcanic eruptions~\cite{Peregrine20}.
 By no means does this short list exhaust the perspectives of social scientists and historians on what drives historical patterns, but it captures many prominent tendencies. How can we unify these perspectives and thereby provide an integrative framework for history? Better still, can we construct such an integrative framework that is well-suited to analyzing the new historical data sets being constructed in addition to older ones now being made publicly available?

Here we argue that we can unify the perspectives presented above by formalizing them all as being simply different aspects of 
stochastic processes that govern the dynamics of human groups through spaces of social, political, economic, environmental, and other variables. In addition, we argue below that due to the relative paucity of sample sizes within historical data compared to the needs of statistical analyses requiring high degrees of freedom, the particular framework of first-order Markov stochastic processes is often well-suited to analyzing historical data sets. 

There are many other benefits to adopting a stochastic process perspective, in addition to unifying the perspectives elaborated by earlier researchers. Most obviously, like any other formal framework, adopting a stochastic process perspective forces us to to be precise and explicit in our thinking about the dynamics underlying any particular historical data set. Furthermore, stochastic process models provide a way to quantify the randomness in the dynamics of human groups, how that randomness varies depending on the characteristics of the group, and how that randomness differs from the observational noise inherent in the construction of all historical data sets. Rather than framing our data around preconceived interpretations, using this framework can allow the data to identify how human social groups evolve, possibly revealing unanticipated relations among the variables in the data.\footnote{Of course, our preconceptions and experience still affect our choice of which variables go into the historical data set in the first place. Our concern here is to able to minimize the role of such preconceptions in the analysis of our data sets, however those data sets were constructed.} 

As a practical matter, many historical and archaeological data sets contain large amounts of missing data. Indeed, one of the major challenges with analyzing archaeological time series data using conventional techniques is the need to either ``imputate'' missing data in the time-series (as in \cite{turchin_seshat_pnas}), or simply remove from the analysis any time-series that has too much missing data (as in \cite{whitehouse2019complex}). As we describe in Section 2\ref{sec:flow} below, using a stochastic process estimator allows us to both avoid the entire issue of imputations, and to use all of our data, without throwing any out.

In addition, having an underlying \textit{first-order Markov} stochastic process model of the dynamics of a social group has particular advantages for the social scientist. Such a model tells us how the future dynamics of a human group directly depends on the current characteristics of that human group, i.e., on the current position in a vector space of such characteristics.\footnote{In contrast, the insights from a high-order time-series analysis of the same data set are often more opaque, being of the form, ``if the social group has had a particular \textit{sequence} of characteristics in a succession of time periods, then its future dynamics is likely to be ...''.} Similarly, it tells us how the randomness in those future dynamics varies directly as one changes those current characteristics. Furthermore, as a practical issue, many historical and archaeological data sets contain very few data in a very large space. This means that a first order Markov model with relatively fewer parameters than higher order models is a sensible baseline model and, further, if the data is particularly sparse it may not even be possible to use a higher order model without over-fitting the data.\footnote{Note that often we will want to have a large number of parameters in our model simply to allow it to capture how the stochastic dynamics varies across the vector space of characteristics of the group. That is just as true for a first-order model as a high-order one.} 


$ $
 
\noindent \textbf{Roadmap:} In the next section we present a high-level summary of stochastic process models, beginning with a discussion of first-order Markov processes, and then discussing higher-order processes. 

To ground these abstract considerations, we then present several examples of how to analyze historical data sets with stochastic processes, chosen to illustrate the strengths of stochastic process modeling described above. The first set of examples presents recent research that explicitly involves stochastic process models. Following that, we present several examples of recent research that do not explicitly invoke stochastic process models, but develop analytical approaches or data sets that (we argue) can contribute to the development of such models. We end with a section suggesting possible near-future research projects fully grounded in a stochastic process perspective.

Importantly, the more powerful the tools for analyzing a data set, the greater the opportunity to misuse them. Therefore estimating an underlying stochastic process from a historical data set has its risks. Accordingly, we also use the examples to illustrate some potential pitfalls and how to avoid them. In the same pragmatic spirit, we present lessons learned in many of the examples. 

Through these examples, we hope to convince the reader that the framework of stochastic processes is particularly well-suited to uncovering deep, overarching regularities in the unfolding of history.
More broadly, we argue that a stochastic process perspective 
provides a way to (at least start to) unify the many perspectives on the unfolding of history promoted by previous researchers.

\section{Stochastic processes}

\subsection{First-order Markov processes}


To make the ideas above more precise, we now introduce the most basic type of stochastic process, which we shall refer to repeatedly in the sections below. This process is called a ``(first-order, time-homogeneous) Markov process'', with the associated equation for the dynamics sometimes referred to as a ``master equation.''

Let $\mathbf{x}$ be the location of a society in some appropriate socio-political-environmental state space, and let $t$ be time. Consider the evolution of the society through that space starting at the value $\mathbf{x}_0$ at time $t_0$. (For the moment we take $\mathbf{x}$ to be a real-valued vector, for expository purposes.)
A fixed (time-invariant) stochastic process assigns a probability to each trajectory of $\mathbf{x}$ emanating from this starting point. 
In a Markov stochastic process, there is a parameter ${\bm{\theta}}$ specifying a ``transition matrix,'' or kernel, $\mathbf{W}_{{\bm{\theta}}}(\mathbf{x'}|\mathbf{x})$, 
and the probability distribution of trajectories through $\mathbf{x}$ is governed by the linear differential equation,
\begin{eqnarray}
\frac{d}{dt} p_t(\mathbf{x}) = \int d\mathbf{x'} \left[\mathbf{W}_{\bm{{\bm{\theta}}}}(\mathbf{x}|\mathbf{x'}) - \mathbf{W}_{{\bm{\theta}}}(\mathbf{x'}|\mathbf{x})\right] p_t(\mathbf{x'})
\label{eq:1}
\end{eqnarray}
Intuitively, the term $\mathbf{W}_{{\bm{\theta}}}(\mathbf{x}|\mathbf{x'})$ captures the flow from all states $\mathbf{x'}$ into $\mathbf{x}$, while the term $\mathbf{W}_{{\bm{\theta}}}(\mathbf{x'}|\mathbf{x})$ captures the flow from $\mathbf{x}$ into all other states. 

As \eqref{eq:1} illustrates, in a stochastic process it is not the state of variable $\mathbf{x}$ that evolves deterministically and continuously. Instead, it is its \textit{probabilities} that do. These probabilities tell us how the trajectory of values $\mathbf{x}(t)$ can fluctuate, in general allowing both smooth, drift-like behavior, as well as discontinuous jumps. 
Importantly, there are two types of such jumps. If the system is currently at $\mathbf{x'}$, and the kernel $\mathbf{W}_{{\bm{\theta}}}(\mathbf{x}|\mathbf{x'})$ places non-infinitesimal weight on an $\mathbf{x}$ that lies a non-infinitesimal
distance from $\mathbf{x'}$, then there is nonzero probability of a discontinuous jump across the space. On the other hand, it could be that a small
change in the {current} position, $\mathbf{x'}$, results in a large change in the kernel $\mathbf{W}_{{\bm{\theta}}}(\mathbf{x}|\mathbf{x'})$, and therefore a large change
in where the system is likely to move from $\mathbf{x'}$. This can be viewed as a ``jump'' in the \textit{derivative} of $\mathbf{x}$, arising from a small
change in $\mathbf{x'}$ 
i.e., a jump the ``direction of flow'' of the system, rather than in the position of the system itself. 
The parameter ${\bm{\theta}}$ parameterizes the stochastic rule, and so directly specifies where in state-space either type of jump occurs.

No matter what the precise stochastic process model being used to explain historical data,
we will refer to the average-case, smooth patterns in the trajectories through the space of random variables as the \textbf{trends} emphasized by structural approaches. These include patterns in the trajectories through the space of environmental variables as well as through more strictly socio-political ones. Similarly, discontinuous \textbf{jumps} in the trajectories of the random variables often but not always involve human agency. The distinction between trends and jumps is often invoked in the historiography of science. For example, the invention of new ideas or technologies can occur through a gradual process, such as with Moore's law. New ideas can also occur through a jump, however, such as Einstein's invention of general relativity.

In addition to such changes in the value of $\mathbf{x}$ at some time $t$ that are governed by the stochastic process, it is possible that the \textit{stochastic process itself} changes at some time $t$. Such a change in the process
would not affect the immediate value $\mathbf{x}(t)$, but rather it 
would affect how $\mathbf{x}$ is likely to evolve after time $t$.\footnote{To be more concrete, suppose that the
stochastic process itself changes at time $t'$. This would mean that if we start the society at $\mathbf{x}_0$ at some time $t_1 > t'$, the likely ensuing trajectories 
would be different from what they would be if we instead start the society at $\mathbf{x}_0$ at a time $t_0 < t'$.}
We refer to such a change in the stochastic process itself as an \textbf{exogenous perturbation} of the stochastic 
process.\footnote{As a technical comment, throughout this paper we view real-world data sets as being formed by noisy observations of trajectories $\mathbf{x}(t)$, rather than incorporating the observational process directly into the stochastic process itself. So we do not consider the possible effects of time-dependence in the map taking $\mathbf{X}(t)$ to our observational data. In particular, for the purposes of this paper, we do not consider the many ways that archaeological data concerning events further in the past can be ``more noisy'' than recent archaeological data.} Formally, an exogenous perturbation is a change at some time $t$ in the parameter ${\bm{\theta}}$ in~\eqref{eq:1}, due to an event not captured in the stochastic process variables, $\mathbf{x}$.


To illustrate these points, suppose $\mathbf{x}$ is a vector of variables related to socio-economic characteristics of a 
particular society. For that space of variables $\mathbf{x}$, slow changes in climate due to reasons independent of human activity
constitute gradual exogenous perturbations. For the same space of variables, events like volcanic eruptions would also be exogenous perturbations, but sudden ones. Similarly, the birth of a great leader who is in fact historically consequential (not just idolized that way by future generations) would be a sudden exogenous perturbation.

It can be quite challenging to ascertain from a given historical data set when (if ever) such exogenous perturbations happened, or if instead the dynamics of the data through the space of values $\mathbf{x}$ just reflects evolution under \eqref{eq:1} 
for an unvarying $\bm{\theta}$. Yet making this determination is crucial if we wish to infer any general principles of historical dynamics. We touch on this issue several times in the examples below, and also discuss it with some actual historical dynamics from the US Southwest in the SI Section 1.A.

The foregoing summary of Markov processes elides many potentially important details. To present some of those details, in the SI Section 1 we work through perhaps the simplest Markov process, a 1-dimensional random walk with step size 1. We also discuss some important variants to \eqref{eq:1} there.

\subsection{Higher-order stochastic processes}


It is important to emphasize that the first-order Markovian dynamics given in~\eqref{eq:1} is only one type of stochastic process. If a system is first-order Markovian then the values of our variables in the past provide no useful information beyond what's in their current value for predicting their future values. 
However, in many common circumstances this property will be violated, at least to a degree, and so the system will not evolve {precisely} according to~\eqref{eq:1}. 

As an example, suppose we coarse-grain the values of $\mathbf{x}$ into large bins, e.g., to help in statistical estimation. Then in general, even if the dynamics over $\mathbf{x}$ is first-order Markovian, the dynamics over those coarse-grained bins will not be. (We illustrate this with the random walk example in the SI Section 1.)
Similarly, if we leave some important components of $\mathbf{x}$ out of our stochastic process model, they become  what are called
``hidden variables''. In such circumstances, again, the dynamics over the visible components of $\mathbf{x}$ will not be first-order Markovian (although it might be $n$-th order Markovian for $n>1$). 

In all of these scenarios, alternative techniques like Hidden Markov models (see Section 2(~\ref{sec:demography})), or high-order Markov models, may be more appropriate than first-order Markov models obeying \eqref{eq:1}. Indeed,
certain types of time-series analysis that have already been used in historical analysis can be formulated as a type of higher order Markov process. For example, the use of delay coordinates \cite{takens1981detecting,sauer1991embedology} to predict future trajectories can be formulated as a technique for estimating the average future trajectory under such a stochastic process.\footnote{Note though that such techniques can also be interpreted as estimating deterministic dynamics embedded on a high dimensional manifold in state-space \cite{takens1981detecting,sauer1991embedology}.} Similarly, vector autoregressive moving-average (VARMA) models \cite{lutkepohl2005new} can also be formulated as stochastic process models. 

While such higher-order models
have been applied before in historical analyses, 
they have several non-trivial shortcomings, as discussed above.
In addition, many of the issues mentioned above (e.g., detecting exogenous perturbations) are generic, in that they also arise in those alternative models. For all these reasons, we mostly concentrate on first-order Markov processes in this paper.

\section{Recent research in history based on stochastic process modeling}
\label{sec:seshat_and_demog}


\subsection{Jumps in sociopolitical complexity of polities in the \textit{longue duree}}
\label{sec:hinge_points}

It is widely agreed that over long-enough periods of time there is a strong trend for polities to increase in size, or to disappear (e.g.,  \cite{tainter1988collapse}). Less clear is whether such increases are relatively smooth trends in which various measures of scale and/or capability, such as information processing or capability at warfare, increase in lock step; or whether there are clear discontinuities or disjunctions in the growth processes. 

The Seshat data set~\cite{francois_etal2016} allows analyses that can address such questions. This data set contains values for a large and diverse set of variables concerning societies over the past several thousand years, in both Afro-Eurasia and the Americas. (We provide more  background in the SI Section 3.) This data set has been condensed for some recent analyses into a set of nine complexity characteristics (CCs) defined by the researchers for multiple past societies. These are reported at (usually) one-hundred-year intervals for the dominant polity controlling each of thirty world regions called Natural Geographic Areas (NGAs; the controlling polity may have its capital outside the NGA). For some NGAs, these time-stamped data stretch back millennia. Principal components analysis (PCA) on the CCs found that the primary component, PC1, explained roughly three-quarters of the variability of the 9 CCs~\cite{turchin_seshat_pnas}. The Seshat team suggests interpreting PC1 as an overall, scalar measure of social complexity\cite{turchin_seshat_pnas,whitehouse2019complex}.\footnote{This is because, by construction, the individual CCs represent separate complexity measures, PC1 ``captures three quarters of the variability''. Note as well that PC1 has a roughly equal loading on all 9 CCs -- that is, it is a weighted version of the individual complexity measures.} Note that PCA is independent of the time-stamps of the individual data points. So mathematically, there is no reason to expect that PC1 bears any relation to time; it could just as readily decrease with time as increase. Nonetheless, examination of the data makes clear that there is a strong tendency for polities (and sequences of polities within an NGA) to increase their PC1 value with time, albeit with long periods of stasis \cite{turchin_seshat_pnas}. Crucially, different NGAs have different starting dates, and there is no guarantee that sociopolitical change will march in lock-stop across NGAs. On the other hand, as discussed below, PC2 consists of two sets of CC components: one related to polity scale and another related to information processing capabilities \cite{Shin2020}. This suggests fitting a very basic ``stochastic process model'' to the Seshat data, by graphing PC2 against PC1, with larger PC1 values generally corresponding to later times in an NGA's sequence.



\begin{figure}
\centering
\includegraphics[width=.8\linewidth]{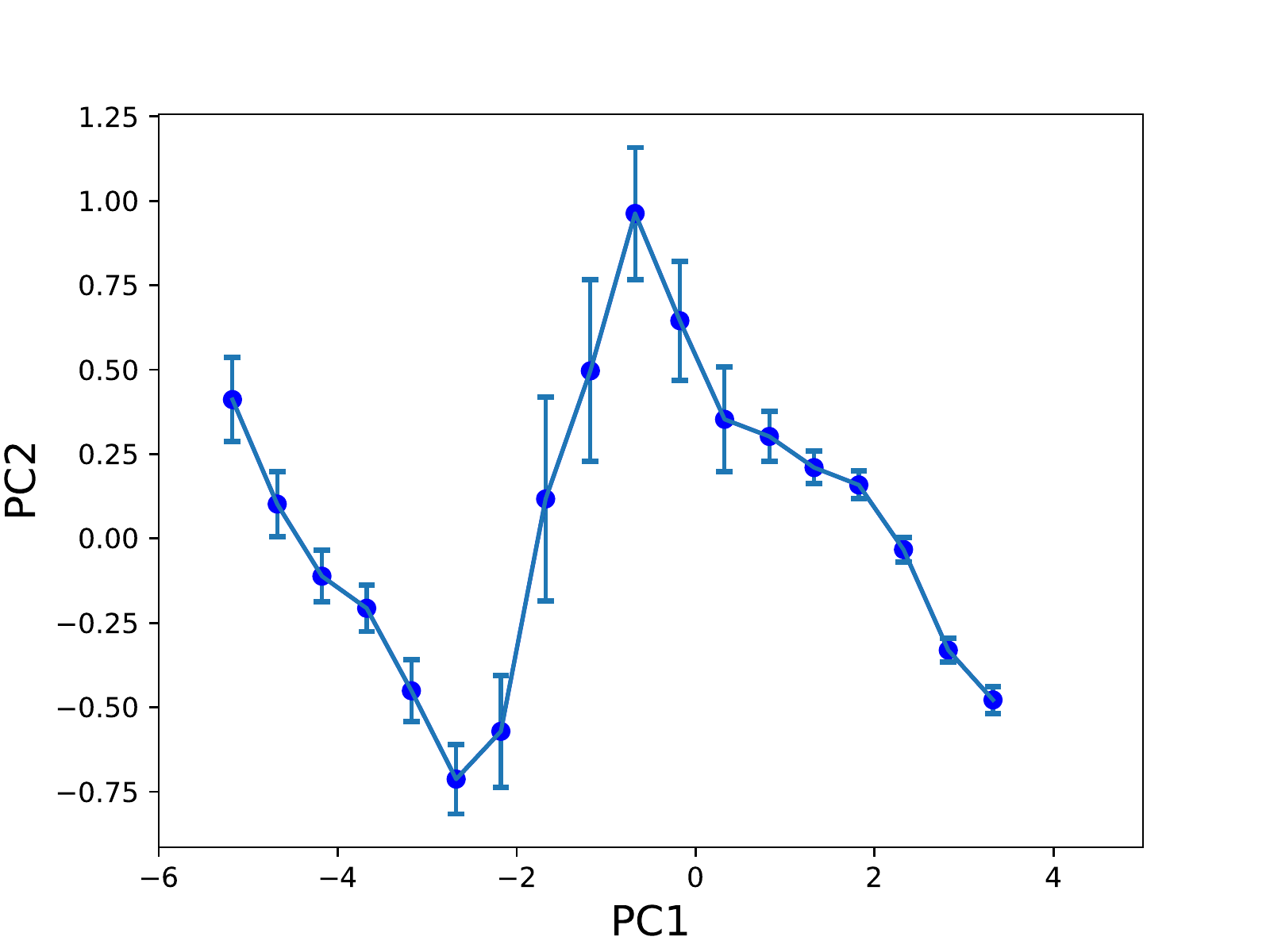}
\caption{The mean value of PC2 as a function of PC1, where the mean and error bars are calculated using a sliding window with a width of 1.0 in PC1-space. This graphic is reproduced from \cite{Shin2020}.}
\label{fig:PC2_vs_PC1}
\end{figure}

A recent paper~\citep{Shin2020} explored this possibility by analyzing how PC2 and PC1 co-vary. Whereas PC1 has positive loadings on all 9 CCs, PC2 has negative loadings on the scale variables (such as polity size) and positive loading on information variables (such as writing, texts, and money). Interestingly, as illustrated in \cref{fig:PC2_vs_PC1}, PC2 is a ``saw-tooth'' function of PC1, first decreasing until PC1 is about $-2.5$, then increasing until PC1 is about $-0.5$, then decreasing again. This reveals an unexpected sociopolitical phenomenon: there is a strong average tendency among the polities in this sample to first increase in size until a certain threshold size is reached (PC1 $= -2.5$), at which point the dynamics is taken over by increases in the information-processing ability of the polity (broadly defined) with relatively little increase in size. Eventually this process reaches a new threshold (PC1 $= -0.5$), after which a dynamic process increasing the size of the polity again takes over.

The key point is that the two values of PC1 at approximately $-2.5$ and $-0.5$ appear to constitute \textit{thresholds in complexity}, with different sociopolitical processes dominating the dynamics depending on whether the PC1 value is less than $ -2.5$, between $ -2.5$ and $ -0.5$, or greater than $ -0.5$. This tendency is exhibited by polities that reached these thresholds at vastly different times, in far-separate locations and differing local environments, with extremely different neighbors. Therefore these two thresholds cannot be due to exogenous perturbations, in which ${\bm{\theta}}$ would change at a specific time. Rather it appears likely that they are examples of jumps of the second type discussed above, where the transition matrix $\mathbf{W}_{\bm{\theta}}(\mathbf{x'}|\mathbf{x})$ in~\eqref{eq:1} changes discontinuously once $\mathbf{x}$ (in this case, PC1-PC2) reaches the associated threshold.

One serendipitous consequence of the sorts of stochastic process modeling we advocate in this paper is that it fosters cumulative cascades of analyses, in which an insight coming out of analyzing one data set allows us to analyze a second data set in a way that would not have been possible otherwise.
We can illustrate this with the current controversy about the relationship between when a polity undergoes a jump in ``social complexity'', and when its members widely adopt worship of ``moralizing gods.'' On side of the debate are researchers who say that the emergence of such gods is a precondition for a jump in social complexity. (Loosely speaking, the reasoning underlying this hypothesis is that worship of such deities fosters wide-spread sacrifice in pursuit of the common good, which in turn results in the jump in social complexity.) Others have argued just as vigorously that the causal influence is the other way around, that such jumps are actually a necessary condition for the emergence of moralizing gods in a society. 

We can exploit our analysis described above concerning hinge points to address this issue. To begin, we suppose
that the first hinge point is a marker for a jump in social complexity. As described in App.\,A(2),
we can then analyze the relationship between the time that moralizing gods appear in a polity and the time when that polity undergoes a jump in social complexity. The results contrast with the somewhat polarized views in the literature ~\cite{purzycki2016moralistic,whitehouse2019complex}. 
Given a null hypothesis that polities with moralizing gods arise later than polities without them, there appears to be little evidence of correlation between the time that moralizing gods appear in a polity and the time that that polity experiences a ``jump of social complexity''.\footnote{We emphasize, though, that we have not yet done a careful analysis of exactly how much correlation there is.} Such gods do not seem to be a precondition for jumps in complexity by a polity, nor does it seem that a polity's having gone through such a jump is an absolutely necessary precondition for their appearance (as claimed in~\cite{whitehouse2019complex}). A natural way to extend this preliminary analysis would be to fit a full stochastic process model to historical data in a state-space $\mathbf{x}$ that consists of both the CCs of a society (or one or more of the corresponding PC-values) and a variable representing the presence / absence of moralizing gods.

\subsection{Are there stable social evolutionary types of societies?}
\label{sec:homogenous}

Papers in the social sciences sometimes note that some set of multiple time-series across a space seem to move quite slowly across one or more regions of that space. This has sometimes been interpreted as meaning that those regions are ``basins of attraction'' or ``attractors'' or ``equilibria'' of an underlying stochastic process  \cite{Peregrine18}. 
A long-running interest in archaeology (stretching back at least to the mid-late nineteenth century \cite{Tylor1871}) is whether there really are such stable social evolutionary types of societies, or whether any apparent instances of such societies in the data are illusory. 

\begin{figure}
\centering
\includegraphics[width=.8\linewidth]{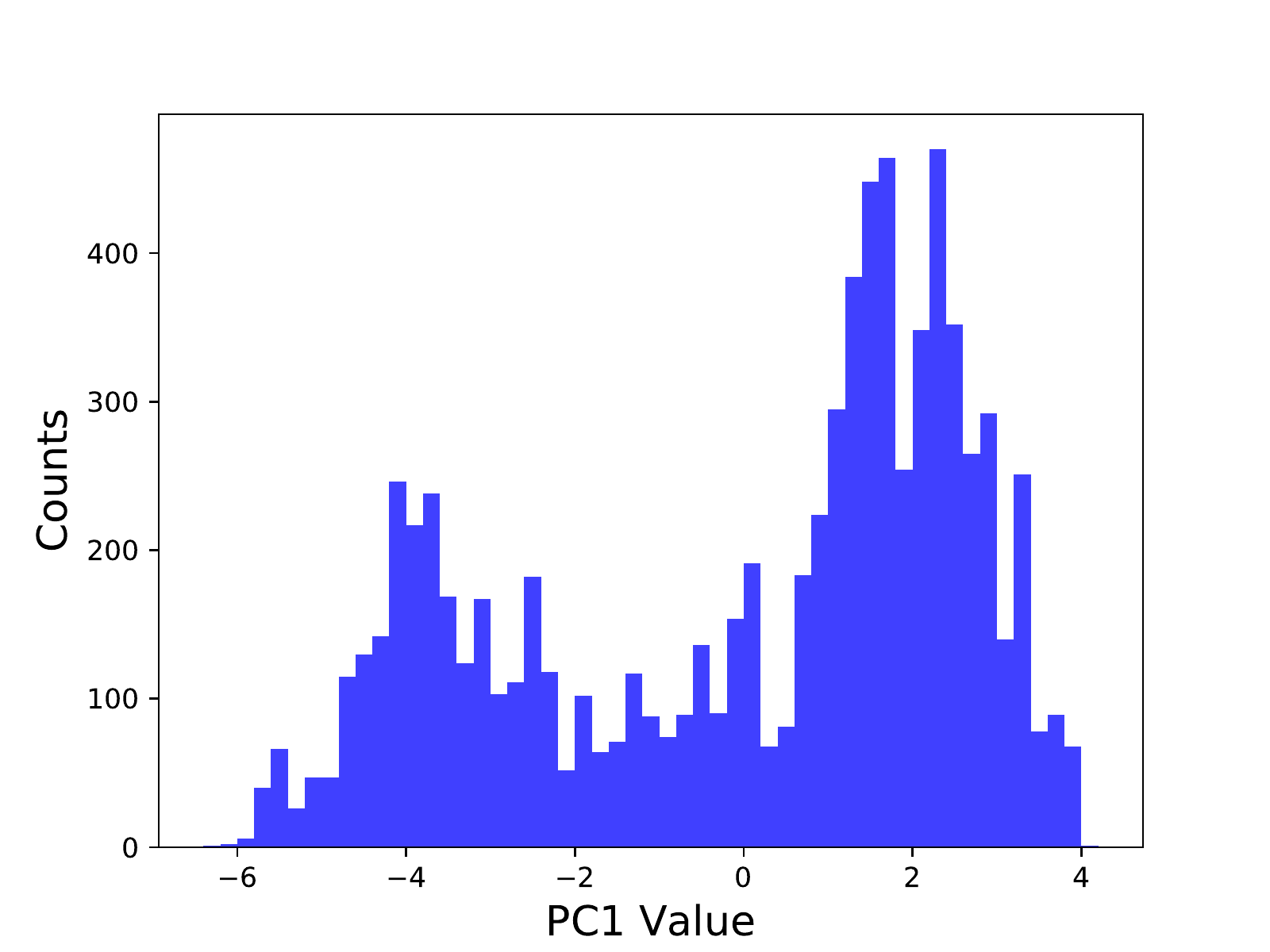}
\caption{Histogram of PC1 values for all polity-time pairs.}
\label{fig:PC1_hist}
\end{figure}

In this subsection we illustrate how careful use of stochastic process analysis can help address this issue, again using the
Seshat dataset \cite{SeshatDB}.
If one plots a histogram of the PC1 values of all the time-stamped observations across NGAs, one finds that, even though every polity is usually sampled once per century (though this is less true early in time for polities with long time sequences), the PC1 values cluster into two widely separated regions (see \cref{fig:PC1_hist}). Does this imply that those regions of PC1 values are ``equilibria'' of the underlying dynamics, i.e., stable social evolutionary types? 

While some have thought so \cite{MirandaFreeman20,turchin_seshat_pnas_SI}, an analysis using the sorts of tools we advocate in this paper suggests that the answer, in point of fact, is no. Those clusters can be explained as arising from a null model that results from the interplay between the starting values of polities in the data set (i.e., the point at which they are first coded), the underlying stochastic process, and the sampling of the stochastic process. To give a simple, counter-intuitive example of that interplay, even if a stochastic process has the same average speed across a space, if the \textit{variance} of its speed varies across that space, then we will see clusters; regions with higher variance will have clusters of more data points than regions of lower variance~\cite{Shin2020}. Complicating the picture still further is the fact that the Seshat data set comprises multiple, different time-series, reflecting a combination of three factors:

\begin{enumerate}
\item [a)] Random variation across the time series of the polities in the PC1 value of the earliest data point;
\item [b)] Random variation across the polities in the chronological time of that first PC1 value;
\item [c)] A general drift of polities from low to high PC1 values that is well-modeled by a time-homogeneous Markov chain.
\end{enumerate}

Perhaps the most conservative way to test whether such factors lead to features like clusters in the data set is to use a null hypothesis that the data were generated by a first-order, time-\textit{homogeneous} discrete-time Markov chain, corrected so that the variance of the chain as one moves across the space matches the variance in the data set as one moves across the space.

Indeed, as described in the supplementary materials of \citep{Shin2020}, one can accurately fit a first-order, time-\textit{homogeneous} discrete-time Markov chain to the Seshat data set PC1 values, mapping one PC1 value to the next with a conditional distribution that does not change in time (as it ought, if polities were ``dwelling'' in certain portions of its space). If one forms 30 sample time-series by running that Markov chain 30 times, each for a different number of iterations, and superimposes the 30 resultant time-series, one finds that they will typically form two widely-separated clusters, just like the clusters in the original Seshat data along the PC1 dimension (\cref{fig:PC1_hist}). Yet because the underlying process is a time-homogeneous Markov chain, there are \textit{no} attractors in the underlying dynamics.~\footnote{A time-homogeneous Markov chain has a unique stationary probability distribution. That distribution can have multiple peaks, but the probability is $0$ of the system staying in one peak indefinitely. In that sense, the dynamics cannot have multiple basins of attraction.}

Intuitively, these clusters reflect several factors. First, the 30 time-series are all transients (none long enough to be samples of the stationary distribution of the Markov chain). Second, they are all of different lengths (due to (a, b) above). Combined with the fact that the generating conditional distribution is non-uniform, the result is that two clusters in the data appear, with no direct significance for interpreting the underlying dynamics. We discussed this example in some detail to highlight the formal and theoretical subtleties in using  stochastic processes to model archaeological data sets. But, as this analysis shows, these challenges can be met---supporting our claim that such a program can now be developed successfully.

\subsection{ Comparing different historical trajectories: Flow maps}
\label{sec:flow}

Another big question in history is to what degree different trajectories are comparable. Formal analysis of the sort presented here allows us to move beyond metaphorical uses of developmental stages or historical cycles. In Sec.\,2\ref{sec:homogenous} we presented a cautionary tale concerning computational history, arguing semi-formally that a simple null-hypothesis test shows that the Seshat data set is consistent with the hypothesis that it was generated by a time-homogeneous Markov chain. While such null-hypothesis tests are a useful staring point, they are blunt instruments that sidestep the critical question of what is the best statistical inference from the data. 

Fortunately, analytical tools developed very recently in other scientific fields might prove quite useful when applied to historical data sets. In this section we illustrate how one such tool can produce a far more sophisticated estimate of the stochastic process that generated the Seshat data than a simple fit with a time-homogeneous Markov chain.

\begin{figure}
\centering
\includegraphics[angle=-90,width=\columnwidth]{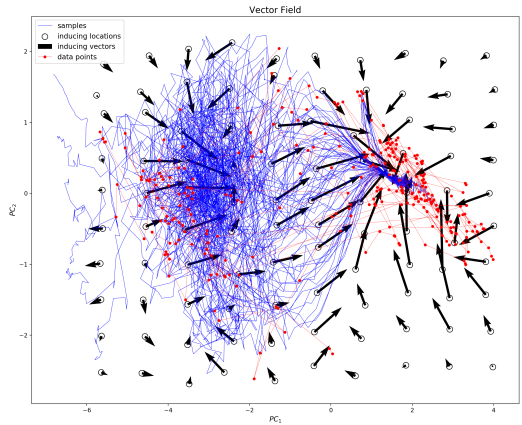}
\caption{The two axes are PC1 and PC2 of the (original) Seshat data set, respectively. The red dots are the elements of that data set. Since those elements of the data set are time-stamped, one can find the Langevin SDE that has highest posterior probability conditioned on that data set. The black arrows are the mean velocity vectors of that Langevin equation at the associated PC1-PC2 positions (i.e., they are values of the drift vector field of that SDE evaluated
on a  grid). Finally, the blue lines are counterfactual, sample trajectories of that SDE.}
\label{fig:sde}
\end{figure}

\cref{fig:sde} 
presents a plot of trajectories of the polities in the Seshat data set in the joint PC1-PC2 space. Each of those 30 time-series is quite short, which makes it
difficult to extend any single one of them, e.g., by using vector autoregressive-moving-average (VARMA) models or delay coordinate techniques
 \cite{lutkepohl2005new}. A different approach is to model the dynamics across PC1-PC2 by fitting the data to a Markov process as in~\eqref{eq:1}, with its
 transition matrix restricted to the set of stochastic processes called ``Langevin equations'' (a.k.a., ``stochastic differential equations'' [SDEs]) where the drift
 and diffusion terms vary across the space. As an illustration, 
\cref{fig:sde} 
shows the drift term of the Langevin equation given by applying the Bayesian estimation technique recently introduced in~\cite{yildiz2018learning} to the time-series across PC1-PC2 in the Seshat data. 

There are several advantages of this kind of fit of an SDE to a set of short time series'.
Most directly, recall that as discussed above, by itself, the fact that PC1 explains most of the variability of the Seshat data set provides no basis for identifying the PC1 position of a polity as a measure of its social complexity, despite the fact that it has often been explicitly used that way \cite{whitehouse2019complex}.
In fact, one could argue that any data set, by itself, cannot provide guidance as to how to quantify ``social complexity''. Ultimately, the problem is that social complexity is a vague, intuitive concept, meaning different things to different people.

In contrast, a flow field like that in \cref{fig:sde} focuses attention on the more concrete issue of how real-world societies actually evolve.
For example, it has long been noted that some sufficiently coarse-grained data sets seem to suggest that there is a rough tendency of polities to go through historical cycles \cite{khaldun2015muqaddimah}. As future work, one can imagine estimating the SDE that generated those data sets, then employing Monte Carlo sampling of that SDE to quantify this tendency of polities to go through such cycles. For example, this could be done by Monte Carlo estimating the probability that a polity currently at a given $\mathbf{x}$ will follow a trajectory that goes at least $\epsilon_1$ away from $\mathbf{x}$ but then returns to within $\epsilon_2 < \epsilon_1$ of $\mathbf{x}$, as a function of $\mathbf{x}, \epsilon_1$ and $\epsilon_2$.
(In this regard, note the suggestion in \cref{fig:sde} of a circulation pattern around the point $(1.5, 0)$.

In addition, it may prove fruitful to re-express the stochastic flow field (or more precisely, the field of drift vectors defining the SDE). In particular, robust algorithms have been developed in computer graphics for inferring a Helmholtz decomposition~\cite{arfken1999mathematical} of such flow fields. Such a decomposition expresses the flow across PC1-PC2 in terms of (the gradient of) a scalar potential field and (the curl of) a vector potential field. The scalar potential field might prove particularly illuminating; level curves of that potential field across PC1-PC2 would be an extremely flexible way of identifying polities all of whom are at the same ``stage of development''. In the absence of a vector potential, the dynamics of polities would simply be gradient descent, descending those level curves. 

At a high level, this joint use of tools for inferring SDEs and the Helmholtz decomposition is a way to infer directly from the data novel \textit{laws} governing social systems (or at least strong empirical regularities emerging from them), formulated in terms of the Helmholtz decomposition. Note though that both the SDEs and the resulting potentials might not have a simple, explicit form. That could make it difficult to give them a simple social science interpretation. This difficulty is actually fundamental in all analysis of longitudinal data sets; in particular a similar interpretational challenge holds even for fitting one-dimensional time series data sets using delay coordinate techniques, even linear ones like ARMA models~\citep{takens1981detecting,sauer1991embedology}; addressing this challenge is a topic for future research. 

\subsection{A hidden Markov model for demographic data}
\label{sec:demography}

While archaeologists often debate which types of variables are most important for understanding a given region's past, a few of these are often considered more crucial than others, including climate (principally temperature and rainfall), demography, political organization, technology, economy, ideology, and local ecology. In this section, we describe in some detail a stochastic process model that can be used for demographic modeling and which can be used to fuse multiple types of data into a single inferential framework. Aside from being potentially useful in its own right, we offer the model as a template for using stochastic process models in archaeological or historical work.

Of the variables related to demography, perhaps the most crucial is a region's total population size, but it is often necessary to consider also how the population is structured by age, sex, space, and socioeconomic status. High-quality demographic models exist for describing structured populations \cite{rogers1966,caswell2001}, so the main challenge is how to use demographic models to infer past demography from the available archaeological data.

One complication is the temporal fidelity at which inference is possible. Seasonal and yearly changes in climate can change demography at yearly and sub-yearly timescales, but it is rarely, if ever, possible to infer both climate variability and demography at these timescales. As described in the introduction, this means that the data is coarse-grained, and so might not be accurately described by a Markov process. Therefore a hidden Markov model (HMM) could prove more suitable. In App.\, D
we describe in detail an age- and sex-structured population projection model. The fundamental demographic variable is the population vector $\mathbf{z}_t$ of females in the population, the elements of which are the number of females in each distinct age class. The time-dependent population matrix projects the population to the next, distinct time-step per

\begin{equation}
	\label{eq:pop_proj_time}
	\mathbf{z}_{t+1} = \mathbf{A}_t \, \mathbf{z}_t \mbox{.}
\end{equation}

\noindent In addition, a corresponding population vector of males exists, which depends on the sex ratio at birth (SRB) of males to females and male age-specific mortality (details in SI), which is rarely equal to female age-specific mortality and may fluctuate more, for reasons like warfare. 

Substantial work in historical demography and life history theory places constraints on the plausible values of $\mathbf{A}_t$ \cite{wood1998,jones2009,jones_tuljapurkar2015}, or (if adopting a Bayesian approach) can be used to specify priors on $\mathbf{A}_t$. To allow for this we assume there exist $n=1,\cdots N$ distinct types of annual, reference population projection matrices. For example, one type could be appropriate during a famine or other stressing event, while another corresponds to a stable population size in which high mortality (especially high juvenile mortality) is balanced by high fertility. Let $\mathbf{p}_t$ be a probability vector of length $N$ that gives the probability of being in a given reference state. Transitions between these hidden states occur per

\begin{equation}
	\label{eq:transitions}
	\mathbf{p}_t = \mathbf{W}_{{\bm{\theta}}} \, \mathbf{p}_{t+1} \mbox{,}
\end{equation}

\noindent where the transition matrix $\mathbf{W}_{{\bm{\theta}}}$ depends on a slowly changing, exogenous climatic variable ${\bm{\theta}}$, which we propose modeling as a binary or categorical variable that only occasionally changes given the fidelity at which climatic reconstruction is possible (hence, the Markovian assumption applies only so long as ${\bm{\theta}}$ is fixed). 

What is appealing about the preceding framework is that it allows a diversity of data to be used to infer $\mathbf{A}_t$ and $\mathbf{z}_t$. This has immense value for two reasons. First, it is usually better to use more data (and never worse, if the statistical model is good). Second, while any given type of data is likely biased in distinct and predictable ways (e.g., skeletal collections usually have too low a proportion of the very young, whereas very old individuals are incorrectly aged), these biases are usually different for each data type and uncorrelated across samples, so inference will be far better when multiple types of data are used and sample sizes increase.

This model could be especially valuable if applied to archaeological case studies for which a diversity of factors and variables have been linked. A pertinent example is the so-called Classic Maya Collapse. Between about AD 750-1000, various parts of the Maya cultural region in Mexico and Central America experienced a breakdown of sociopolitical systems. Associated with this were major demographic changes (migrations and, likely, overall population decline), changes in long-distance trade routes, intra-elite competition, warfare, and environmental degradation, all in the context of severe drought \cite{webster2002,aimers2007,kennett_etal2012,turner_sabloff2012,hoggarthetal17}. Exactly how these factors are linked remains an open question, yet a stochastic process model would likely help make these linkages explicit were it to be correctly fitted to the data. For example, drought has been proposed as the major, causal factor leading to the breakdown of sociopolitical systems, but others argue that all these factors, and more, played some role; using our proposed modeling framework would help elucidate the causes.

\section{Previous studies recast in terms of stochastic process models}

\subsection{Integrating Different Data Types for New Insights}

Like demographic processes, environmental processes have long been seen as catalysts of past social change. In particular, environmental processes are central to research on the history of human-centered food webs  \cite{crabtree_reconstructing_2017, DunneFoodWebs2016}. The benefits of bringing a stochastic process perspective to this research can be illustrated with the investigation in Crabtree et al. \cite{crabtree2019subsistence}. That research focused on the question of why small-bodied mammals went extinct in portions of Australia \textit{after} people were removed from the Western Desert. 

In the absence of humans, animal populations are often modeled as evolving via Markovian processes  \cite{Meyn_markov_chain_textbook}. However, \cite{crabtree2019subsistence} demonstrated that the extinctions experienced in the Western Desert would not be predicted from the natural baseline extinction rates of such a process. Adopting the perspective of this paper, this result highlights an important modeling issue: did the extinction of small-bodied mammals reflect a change in a hidden variable of an underlying Markovian process, or was it due to a change of a parameter of such a process, via an exogenous perturbation? This distinction is critical because there are statistical techniques for predicting the evolution of Markovian processes with hidden variables, such as time-series analysis, described above. It might be possible with such a model to predict a change in the dynamics of animal populations by looking at time-series data of their dynamics. However, by definition, if the change in the dynamics was due to an exogenous perturbation, then it cannot be predicted from any time-series data, no matter how exhaustive.

In this particular case, it seems unlikely that the change in the dynamics of the animal population level $\mathbf{x}$ can be accurately modeled as the change in the value of a hidden variable, e.g., by using a time-series model of the dynamics of $\mathbf{x}$. In other words, it seems likely that the change in the dynamics of the animal populations reflects an exogenous perturbation of the parameter ${\bm{\theta}}$ governing those dynamics. However, this has yet to be formally established. In addition, a more elaborate stochastic process model might expand $\mathbf{x}$ to include some variables concerning the people who eventually forced aboriginal peoples to leave. In such a model, changes to the dynamics of animal populations might reflect the ``second type of jump'', described above in the discussion just below \eqref{eq:1}. If that were determined to be the case, it would mean that future values of $\mathbf{x}$ might be predictable, and therefore in particular future values of the animal population level might be predictable. 

As another example, Yeakel et al. \cite{Yeakel_egypt} leveraged Egyptian art by coding the taxa present from the Narmur Palette and other datable art work to model animal species extinctions over time. By combining these data with contemporaneous climate data, they found that aridification pulses in the context of a growing Egyptian population played an important role in destabilizing the ecosystem. In the language of stochastic processes, if we take $\mathbf{x}$ to be the combination of human and animal population levels, we see a trend as growing human populations exerting increasing pressures on the ecosystem. These pressures were exacerbated by exogenous perturbations in the form of a series of aridification pulses and a worsening climate. 

This example suggests that it may be possible to look at the dynamics of human and animal populations in other regions and infer when aridification pulses, or indeed other climatic events, likely occurred, by testing for exogenous perturbations to the trends in the dynamics of the joint human / animal population levels. In the original context of ancient Egypt, where we already have identified \textit{some} exogenous perturbations in the form of aridification pulses, we could perhaps test for the presence of \textit{other} exogenous perturbations, e.g., due to invasions, or perhaps even due to causes currently absent from the historical record. More broadly,  a stochastic process framework could suggest a new avenue for research for Egyptologists: the impersonal trends of extinction and exogenous perturbations of aridification likely had feedbacks on Egyptian society, from impacts on hunting to impacts on cosmology. Exploring how the societal impacts at a given time varied depending on whether there was only an underlying trend  or also an exogenous perturbation at that time could lead to new insights on the dynamics of Egyptian social structures.

\subsection{Stochastic processes over the structure of language and over how language is used}
\label{sec:language}

Stochastic processes that are themselves superpositions of Markovian time evolution and branching processes describing the temporal evolution of features in systems that from time to time break up into parts that evolve (largely) independently thereafter. Such processes underlie the history of biology and human languages and are the base of phylogenetics. Data on cognates across many languages, together with a small number of historical anchor points, allows for the reconstruction of language trees and assign approximate dates of divergence. A careful, quantitative analysis of Polynesian languages, for instance, revealed some periods of ``stasis'' in the history of subfamilies that correspond to distinct phases in the settlement history of the Pacific Islands \cite{Gray:11}. Similar arguments have shed light on the expansion of Indo-European language families. Analogous reconstructions of historical population distributions have been conducted using human genetic markers. The idea to integrate genetic, linguistic, and archaeological data is of course not new \cite{Scheinfeldt:10}, although so far this has been done at the level of interpreting the data rather then integrating them into a common process model. The latter would be desirable in particular because linguistic changes are strongly impacted by contact phenomena such as borrowing, and even correlate with extra-linguistic factors such as the prevalence of agriculturally produced food \cite{Blasi:19}. At present, quantitative studies in linguistics are almost always based on data sets that are the result of extensive manual curation --- although modern methods in natural language processing might be suitable or at least adaptable to tap into much larger resources \cite{Bhattacharya:18a}.

Modern large-scale digitized historical archives often provide high-resolution data on the words that individuals---often, though not always, members of a polity's elite---were writing and sharing with each other. Tools from natural language processing are now both simple enough and robust enough to allow social and political scientists to track the flow of complex patterns of language use that correspond to concepts and habits of thought. 

Ref.~\cite{barron2018individuals}, for example, used the French Revolution Digital Archive (\url{https://frda.stanford.edu}) to show the different roles that members of the French revolutionary parliament played in introducing, sustaining, or rejecting novel ideas in the speeches of that country's constitutional debate. These ideas were discovered in an ``unsupervised'' manner; rather than pre-determining a list of important ideas, and words that corresponded to them, topic modeling automatically extracted word patterns that could then be back-validated on the speeches themselves. Discovered topics include, for example, concepts as fine-grained as ``the possibility of enemies of the revolution within the military''.

\subsection{Analyzing the Great Acceleration}

We are currently living in the Anthropocene. One of the more striking characteristics of this period has been the exponential growth in a large number of important metrics in earth and social systems since the beginning of the 1950s. These growth dynamics are often collectively referred to as the ``Great Acceleration'' \cite{Steffenetal04}. What \textit{caused} these transformations is less clear however.  

All the metrics that have contributed to the Great Acceleration are, however, dependent on population size, which itself has been changing during this period, to no small degree as a consequence of growth in these metrics. This suggests we fit our data concerning those factors with a stochastic process model, to gain quantitative insight into their joint evolution (see appendices for examples of such a model). In particular, because of the specific growth patterns of the variables, it is natural to fit the data to a stochastic process over the \textit{logarithms} of the variables. When applied to historical data sets, this variant of stochastic process models is known as \textbf{historical} or  \textbf{temporal scaling analysis}. Not only can it give us insights into the joint dynamics of population size and the metrics considered by researchers of the great acceleration. In addition, when performed for different temporal intervals, it gives us insight into \textit{changes} of those joint dynamics. This then invites us to investigate whether those changes in the dynamics are due to exogenous perturbations and /or due to changes in hidden variables (as in the analysis of HMMs in Sec.\,2\ref{sec:demography} and / or due to the system reaching a new point of its state space (as in the case of hinge points, considered above in Sec.\,2\ref{sec:hinge_points}).

To make this more concrete, suppose that we find that the scaling coefficient remains stable across time. That would be evidence for a simple trend of the underlying stochastic process. Suppose, as an alternative, that the scaling coefficient changes over time. This is evidence for one of three phenomena: either a concurrent change in some exogenous factor driving the dynamics (like a change in climate, or a major drought, or the Green revolution of agriculture); a concurrent change in a hidden socio-political variable generating the observed historical dynamic (like a major war, a world-wide depression, or changes in financial regulation) or simply the system reaching a critical threshold --- a new part of the socio-political space --- in which the dynamics are different. This illustrates how the stochastic process perspective could invite a host of new investigations, if it was found that the scaling coefficients changed over time.

A recent analysis of the dynamics of the Great Acceleration \cite{PainterDeryc20} we determined that socioeconomic, technological, information, biological and earth systems, when scaled to population size, sort into four distinct growth patterns. These patterns suggest fundamental differences in the underlying network of interactions causing the observed patterns of change. Besides the already known sub-linear, linear and super-linear patterns we detected a novel fourth pattern for the Great Acceleration, 
with scaling coefficients larger than 3, which is unusually large. This pattern applies to parameters that are no longer limited by person to person interactions such as those related to knowledge, technology and finance. These parameters can be interpreted as the main drivers behind the ever accelerating growth dynamics of the Great Acceleration. Decade by decade comparison of these parameters also revealed patterns of change during this time period that correspond well with the possible sources of change that can influence the dynamics of an underlying stochastic process. 



Whether an event or a discovery (such as those contributing to the Great Acceleration) is considered a transformative innovation or just a random occurrence affecting a stochastic process can often only be decided in hindsight. What can be done, however, is to asses to what extent such events were surprising or predictable, given the context of the times--here represented by an underlying stochastic process model. Emerging machine learning methods, especially deep learning neural networks, have succeeded at dramatically improving the prediction of quantities across a wide range of contexts. These models have begun to enter the social and historical sciences as complex event and trend prediction \cite{bainbridge1995neural, davidson2017black, carley1996artificial, zeng1999prediction, maltseva2016data, zhan2018small, macleod2016identifying, bacak2019principled}. We can apply this methodology to a number of different events, such as a regime change, adoption of a  technology or the emergence of an institution with a particular function (e.g., urban garbage collection). 
Prediction targets can also be more complex and specific. It has been applied to predict the location, amount, denomination, and material composition of an archaeological cache of money and precious metal hordes, and the distribution of mints from which the coinage derives. This might only be limited by the area, such as surrounding the Mediterranean, and a period of time to which the hordes date (see appendices for a more in depth discussion of these methods).

\section{Future work and ``aspirational'' case studies}

\subsection{Combining time-series' of hundreds of random variables}
\label{sec:combining_time_series}

As far more data sets become available we will be confronted with combining multiple time series of systems evolving from up to hundreds of different variables, including not only the sorts of variables recorded in Seshat, but also completely different kinds of variables, like lead levels in glaciers, or tree ring widths as climate proxies, to large-scale characteristics of polities that others have discussed before. We will be combining these with wholly different kind of time series, including things like word usage patterns in historical documents, structures in legal codes, time series of fashions, ecosystem characteristics, etc., all concerning different random variables. 

As always, great insights will accrue from fine-grained analysis, in which domain experts deeply scrutinize only a few of these time series at once. 
However, there is also the great allure of integrating all of these myriad time series in a systematic, statistically principled way, to uncover unanticipated connections and insights. Indeed, ultimately, one would want to be able
to integrate all of these time series into a single underlying stochastic process, with associated error bars. However, there are many analytical challenges to doing that, due to the sheer number of these times series, and the sheer breadth of the types of associated random variables. 

As an important example of such a challenge, how does one infer causalities in a statistically principled way among hundreds of time series, all involving different random variables? For example, what techniques would allow us to uncover statistically significant causal connections relating time series ranging over hundreds of different spaces? Can we scale up techniques like Granger causality from econometrics \cite{Granger1969} or transfer entropy and directed information from information theory, to such large numbers of spaces \cite{schreiber,directed_info}? 

As another example, can we extend breakpoint analysis / change detection to involve multiple time series, in order to find non-stationary breaks in the underlying dynamic process? Such breaks would reflect some exogenous events, perturbing the underlying dynamics of the system. So, for example, such a breakpoint might indicate that a particular leader of a state was indeed consequential, rather than just being a ``product of their time", in that they perturbed the underlying dynamics of sociopolitical processes --- we might conclude from such a breakpoint that a leader had literally changed the course of history.

\subsection{Leveraging agent-based models for stochastic process modeling}

Agent-based models (ABMs) are becoming increasingly instrumental for investigating processes that leave limited physical traces in historical and archaeological data, such as work examining the dispersal of hominins out of Africa \cite{Romanowska_movius_2017}. 
Typically such models involve the dynamics of variables that are hidden, 
in that they do not directly appear in the historical data.
For example, Crabtree and colleagues \cite{crabtreeetal17} and Kohler and colleagues \cite{Kohleretal18} built a model for sociopolitical evolution in the North American Southwest in which territorial groups undergoing population growth may succeed in subjugating or merging with other groups via warfare or intimidation. Networks of flows of maize as tribute are among the many outputs from these models. These flows constitute hidden variables have not been observed directly in the archaeological record. 

Adopting this perspective on ABMs suggests many new ways they can be combined with stochastic process modeling. Most obviously, if our original data set has lots of missing values, by fitting the parameters of an
ABM to match the data we \textit{do} have, we could use the values the ABM assigns to the variables with missing data as estimates of those missing data values. In this sense, ABMs can sometimes be used as a variant of the technique of imputations, used so extensively to deal with missing data in the original Seshat analysis~\cite{turchin_seshat_pnas}. As another example, we could sample the values of the hidden variables in the ABM model at regular time-intervals. By adding those samples to the original data set, we could produce an estimated data set in a much larger space than the original data set. For example, if we fit the parameters of an ABM model so that it reproduced the data in Seshat, then we could sample the variables of the ABM model at single century-intervals, and add those sampled values to Seshat, to produce a dataset in a much larger space. We could then perform any of the stochastic process analyses described above to this augmented data set. For example, we could perform PCA on this augmented data set, and examine how the PC2 of this new PCA varies as the new PC1 increases, perhaps finding a more elaborate version of the hinge points that were found
by analyzing the original Seshat data set.



\section*{Conclusion} 

The concept of history unfolding stochastically is not new; in the context of the history of life on Earth, Stephen J. Gould famously asked what would happen if we could ``replay the tape'', which implicitly supposes that an underlying stochastic process generated that tape \cite{Gould89}. Similarly, stochastic process modeling of environmental dynamics has been used to infer exogenous perturbations to the dynamics of human social systems \cite{malik2020uncovering}. In addition, the phylogenetic tree reconstructions of human language dynamics discussed in Sec.\,3\ref{sec:language} have
been used as a ``clock'' to infer the dynamics of socio-political phenomena \cite{currie2010rise,sheehan2018coevolution}. There has also been some 
work directly applying time-series analysis techniques to socio-political datasets \cite{turchin2018fitting}.

These are isolated instances though, rather than a systematic scientific program.
Here we propose something more fundamental: that by grounding our investigations of human social dynamics in stochastic process models, we can not
only better investigate the historical record, but also begin to unify the myriad approaches that have been championed for analyzing that record. Such a program would also potentially allow us to detect drivers for the historical processes that generated that historical record --- in particular, drivers that had not already been anticipated in social science models. This might allow the data to drive our formulation of social science models, as an adjunct to  the more conventional approach under which we analyze datasets only after we first formulate models (e.g., based on intuitive insight and / or on analogizing with models from other scientific fields). Crucially, as we illustrated with the examples above,  both the data sets and computational tools necessary for this vision to become a reality are now coming into being.

It is important to emphasize that we do \textit{not} argue that one specific stochastic process we have identified generates the dynamics of history.   (Indeed, we expect that it will be most fruitful to view history as multiple, interwoven stochastic processes, all with different characteristics.) We are not even advocating whether a time-homogeneous process or time-inhomogeneous process be considered. Ultimately, as in all statistical analysis, the choice of model to fit to the data is governed by considerations of number of data, size of the space of variables, types of variables, etc., with cross-validation used to help winnow the options. (See SI Section IA.) 

We are also not advocating that one specific state space be used to model the stochastic process(es) of human history. Nor are we arguing which subsequent analyses should be applied to stochastic process(es) models inferred from historical data, e.g., to uncover possible causal relationships among historical variables. 

More generally, we are also \textit{not} arguing that all of historical analysis should be formulated in terms of stochastic processes. Statics, as opposed to dynamics, is also (obviously) an extremely important aspect of historical analysis, as history is not only concerned with time series analysis, but also with revealing the internal structures of societies and the patterns of their interactions at any given single point of time. Indeed, even in those sciences where all phenomena are based on a single dynamic law, like quantum physics (Schr\"odinger's equation), much research focuses on statics rather than dynamics. 

Finally, we note that a stochastic process formulation is also central to the other historical sciences, ranging from biology to meteorology to geology. So not only does this perspective allow us to unify the analyses of computational history, it also allows us to align how we investigate human history with how it is done in the other historical sciences.

$ $

\noindent \textbf{Acknowledgments} We thank S. DeDeo for useful conversations in the early stages of formulating this paper.
This paper developed out of a Working Group convened by the Santa Fe Institute in January 2020, supported in part by the National Science Foundation under Grant No. SMA-1620462 to D.H.W and T.A.K.
%
%

\newpage



\clearpage

\appendix

\section{Types of dynamics}

\subsection{Which stochastic process model to use}



Although we believe their are many advantages to using stochastic processes, there are also potential limitations and important considerations in choosing the type of stochastic process model to use. In this appendix, we discuss some of these issues, and provide further details on examples discussed in the main text. In many cases, the examples offer a concrete illustration of the limitations and pitfalls that we must otherwise discuss fairly abstractly (and, of course, of the benefits). One major set of issues has to do with the sparsity of archaeological data, which means that a formal stochastic model may not capture every salient aspect of the socio-political-environmental dynamics, and implies that, a \emph{priori}, we should not necessarily propose detailed models with too many parameters and explicit features. This is one benefit using first order Markov models rather than more complex models with more delays. This problem can be somewhat mitigated by using prior information in a Bayesian modeling framework. 

We should regardless be aware of the shortcomings of first order Markov models. We treat potentially deterministic fluctuations whose underlying causes we do not grasp as stochastic. Some models may be ``blind'' to details of human agency. Since the data and models operate at a rather coarse level, as will be discussed below, it is possible to violate the Markovian assumption. Similarly, even though the some underlying variables, such as humans, polities, dollars, etc., are discrete, we often work with continuous state spaces to simplify the mathematics. Fortunately, this is usually a good and legitimate approximation. In spite of these and other shortcomings, these first order Markov models offer crucial advantages, as they allow us to capture dominant features  and qualitative aspects in a robust manner. They can be flexibly adapted to add complexity when new data come in. They are simple enough for understanding the dynamics they generate.  They may allow us to identify driving forces and critical turning points in historical processes.

There are many variants of the basic first-order Markov process given in
Eq. 1. For example, if the state space is discrete rather than real-valued, then the integral in
Eq. 1 gets replaced by a sum. If in addition one models a historical process as evolving in discrete time, e.g., years, then the derivative on the LHS of 
Eq. 1 gets replaced by a discrete-difference. In fact, that is also the setting of the formal example discussed below,
as well as the \textit{discrete-time} Markov chains discussed in Section 2B and Section 2D of the main text.

It is important to note though that there are subtle assumptions that arise if we use a discrete-time Markov chain. It turns out that a sizeable portion of all discrete-time Markov chains are theoretically impossible, if one presumes that the true underlying Markov process is actually continuous in time.
As a striking example, suppose we have a system with only two states, $\{-1, 1\}$, e.g., due to coarse-graining. The simple discrete-time Markov chain that flips the two possible states, sending $-1 \leftrightarrow +1$, cannot even be approximated as arising from an underlying continuous-time Markov process over those two states~\cite{owen_number_2018,wolpert_spacetime_2019}. In fact, the set of all discrete-time Markov chains
that cannot even be approximated with a continuous-time Markov process has nonzero measure (according to any of the usual measures over the space of stochastic matrices defining discrete-time Markov chains). The same is true
if we restrict attention to discrete-time Markov chains that (unlike the bit flip) are highly non-deterministic.    
Since the physical world is in fact continuous in time, this means that if one wishes to fit a discrete-time Markov model to time series generated by some evolving physical system --- including sociopolitical systems --- one should exercise great care, to avoid accidentally selecting a Markov chain that is physically impossible.

In addition, even in the context of continuous-time models, the assumption of a first-order Markov process is a very strong one. Formally, it means that knowledge of the current state of $\mathbf{x}$ suffices to compute the probabilities of its future states. This amounts to assuming that due to the nature of the variables in $\mathbf{x}$, knowledge of past values of $\mathbf{x}$ will not lead to better predictions of future values, beyond knowing the current value. 
This assumption is satisfied when all relevant details of the state are known. 
However, if we coarse-grain the state, that is, lump similar values together into some kind of coarse or macro state, then the dynamics of those states need no longer be Markovian. (See \cite{Pfante14} for a systematic analysis of coarse graining.) Likewise, if there are hidden variables that cannot be directly observed, but influence the dynamics of the observable state $\mathbf{x}$, the latter's dynamics need not be Markovian. Generally, when our information about the current state is incomplete, be it due to coarse graining, hidden variables, or some other factor, we may have to draw upon the memory of the states $\mathbf{x}$ to make better predictions. 

Ideally, we could do this by fitting a higher-order Markov model to the data. Such an approach is closely related to delay-embedding
techniques~\cite{bradley-kantz-nonlinear-time-series}. (Note that delay embeddings capture chaotic dynamics, which is not possible with first-order
Markov processes.)
However, in practice, the amount of data one needs to fit an order-$n$ model grows exponentially with the 
the size of the space and the value of $n$ --- using a larger value of $n$ will result in a poor statistical fit. Especially in the context of fitting historical data-sets, where data is quite sparse, this can mean that for purely statistical reasons we have to either choose $n = 1$, or adopt a careful Bayesian analysis if we wish to fit the data with an $n > 1$ model. 
(However, see~\cite{malik2020uncovering} for a recent example of trying to fit higher-order models with non-Bayesian methods even when data are sparse, in the specific context of historical data.) 

In practice, it is probably most common to use cross-validation to determine $n$, as well the other hyperparameters in one's model, even if one adopts a Bayesian approach. It is worth noting that there are alternative approaches though, which don't involve cross-validation.
For example, in a hierarchical Bayesian approach, one would average over the hyperparameters according to a hyperprior.
As another example, one could set hyperparameters using the semi-Bayesian approach of ML-II. (As a technical comment, the use of Bayesian ``Occam factors'' should not be used to choose $n$, since they build in a bias to low-dimension models; see \cite{wolpert1995bayesian}.)

\subsection{Noise versus chaos versus bifurcations}

Often stochastic processes can be viewed as a deterministic evolution of the variable $\mathbf{x}$ with noise superimposed. In particular, for a broad class of functions $\mathbf{W}_{\bm{{\bm{\theta}}}}(\mathbf{x}|\mathbf{x'})$, Eq.\,1 in the main text, which involves the dynamics of a time-dependent probability distribution $p_t(\mathbf{x})$,
can be reformulated as a noisy equation for the time-dependent state of the system, $\mathbf{x(t)}$:
\begin{eqnarray}
\frac{d}{dt} \mathbf{x(t)} = A(\mathbf{x(t), t}) + B(\mathbf{x(t), t}) \xi(t)
\end{eqnarray}
where the $\xi(t)$ is the Wiener noise process, and the functions $A(.)$ and $B(.)$ are determined by $\mathbf{W}_{\bm{{\bm{\theta}}}}(\mathbf{x}|\mathbf{x'})$.
(This is the ``Langevin equation'', discussed in the text.) $A(\mathbf{x(t)})$ can be viewed
as the deterministic dynamics of the variable $\mathbf{x(t)}$, with $B(\mathbf{x(t)})$ determining the amount of noise superimposed on that dynamics.


The conceptual distinction between deterministic and stochastic dynamics can get blurred in practice because the deterministic dynamics may be chaotic and therefore appear random. Chaos essentially means that very small fluctuations can get amplified so that from very similar initial conditions very different endpoints can be reached after a long enough time, even if the dynamics are deterministic. Fortunately, people have developed sophisticated methods to distinguish chaotic and stochastic components in time series \cite{kantz1997}. The formal concept of a stochastic process can accommodate both deterministic and stochastic features. In such a process, the probabilities for a state variable $\mathbf{x}$ change in time  according to some rule that is described by some parameter $\theta$. The state $\mathbf{x}$ is observed, while the parameter $\theta$ defines the model and can only be statistically inferred, but not directly observed. $\theta$ itself may also change, in which case the process is called non-stationary, but usually on a slower time scale than $\mathbf{x}$. In nonlinear dynamics, the qualitative properties of the dynamics may change at particular values of $\theta$. One speaks of bifurcations. That is, a very small variation of the parameter can send the dynamics into completely different regimes. Thus, while in chaotic dynamics, the future of a trajectory may depend very sensitively on the initial conditions, at bifurcations, the dynamics depends very sensitively on a parameter value. It is clearly important to identify such bifurcation points in historical dynamics.


\section{Previous examples considering history as a dynamic process}

\begin{figure}
\centering
\includegraphics[width=.9\linewidth]{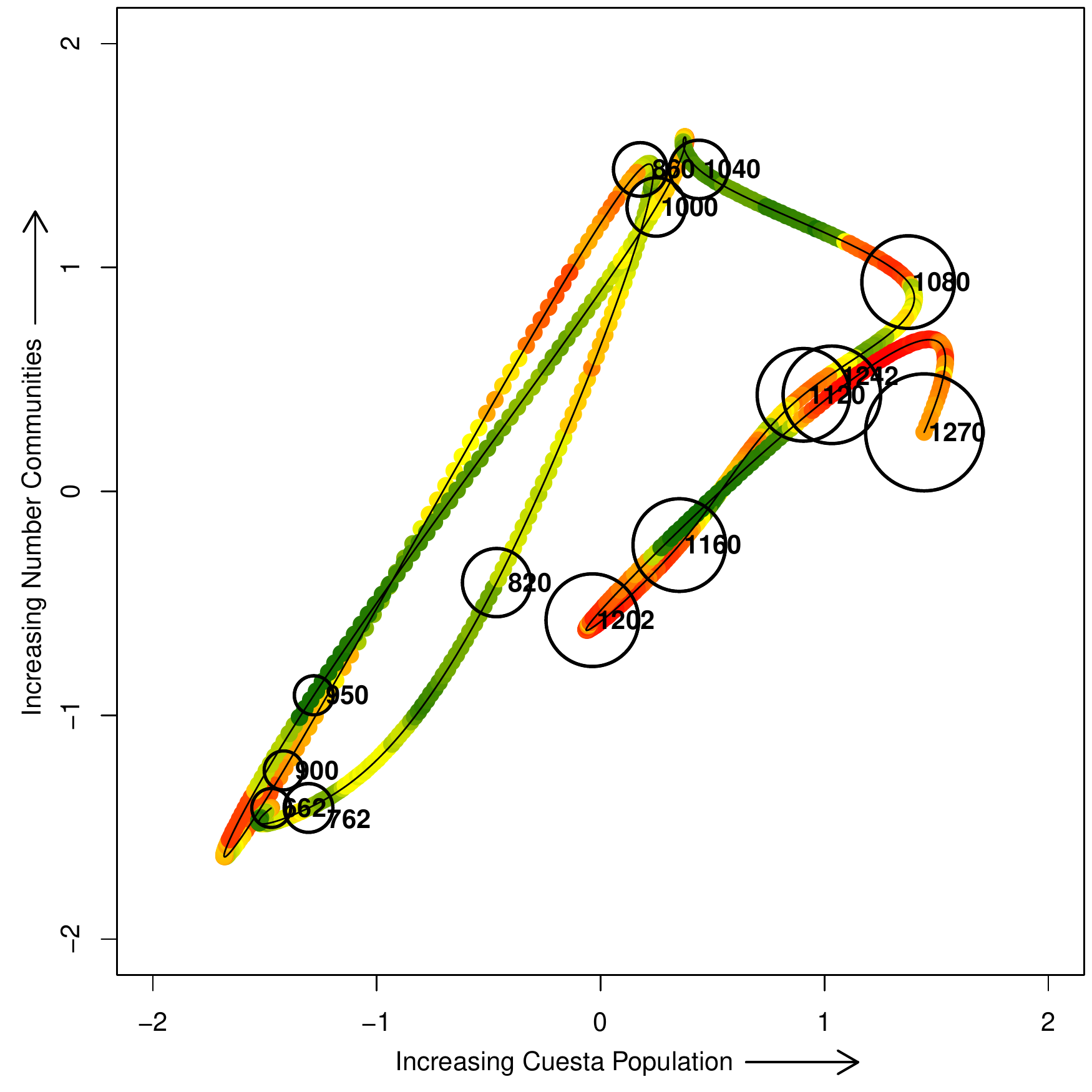}
\caption{Relationship through time between number of communities and  population on the Mesa Verde cuesta, Colorado shown in z-score space across the periods used by the Village Ecodynamics Project. This area was first densely colonized by farming populations ca. A.D. 600 (at bottom left of figure, unlabeled), and was completely depopulated by ca. AD. 1285. Each point is plotted at the midpoint (AD) of each of the 14 periods recognized; the size of each bubble is proportional to average community population at that time. A 20-year smoothed maize productivity niche is shown on a red (low productivity) to green (high productivity) spectrum.  Source: \cite[Figure 6]{Reeseetal19}}. 
\label{fig:reese}
\end{figure}

For some time a few archaeologists and historians have been graphing behaviors of societies through time in small state spaces, usually considering two variables at a time.  Such phase plots implement part of the program we discuss here, since they make it possible to describe trajectories through time in these small state spaces, though they do not generally attempt the fundamental step of formulating the stochastic process model that underlies the behaviors such plots reveal. They are descriptive, graphic devices that do serve to identify semi-cyclic tendencies and possible discontinuities through time. Examples include \cite{KohlerCC09}, who examine the frequency of interpersonal violence against population size through time, and \cite{Reeseetal19}, in which the number of communities and the population size of a study area are plotted against each other, using line colors and symbol sizes to put these into the contexts of estimated maize production levels and average community sizes (\cref{fig:reese}). In the case considered in that figure, a population bearing a new sociopolitical system intruded on this area between the AD 1040 and 1080. From the perspective of this study area, that immigration can be considered as an exogenous perturbation that changed the subsequent evolution of the social and settlement system by (among other things) allowing larger communities to be supported (hence, changing $\theta$). 

Peter Turchin and colleagues have investigated the dynamics of agrarian states with special attention to their demography, social structure, surplus extraction, and sociopolitical instability \cite{TurchinNefedov09}. These variables, and their proxies, are expected to exhibit a patterned relationship with each other through time based on theory developed in \cite{Goldstone91}, \cite{Turchin03}. These examples go further than the archaeological cases mentioned above by positing formal models, though there is no attempt to quantify the fit between any of these models and the empirical examples, or to derive models directly from the data. 

\section{Using PC1/PC2 hinges as reference points for the appearance of moralizing gods}

So long as one can evaluate the value of Seshat's PC1 for the polities recorded in that data set, one can quantify how ``complex'' those polities were when they underwent events recorded in that data set. In earlier work with Seshat this was done simply by identifying PC1 with complexity. In particular, \cite{whitehouse2019complex} used a time-series fit to Seshat data to determine when ``jumps'' occurred in the complexity of individual polities.
The times of those jumps were then compared to the time of appearance of Moralizing Gods (MGs) in the polity.\footnote{Presence of moralizing gods has now been added as variable in Seshat, to allow this analysis by \cite{whitehouse2019complex}, but was not a component of the PC values discussed in this section.} The conclusion was that jumps in complexity are preconditions for the appearance of MGs.

Note that this approach assumes that all intervals in PC1 space correspond to the same amount of ``social complexity'',
since it identifies large changes in PC1 during small times as ``jumps in complexity''.
In addition, in order to assign a time of the complexity jump to any particular polity based on its time-series, which is then used to determine
whether the jump was before the MG onset for that polity,
any time-series was discarded from the data set unless it had a continuous sequence of PC1 values stretching to before the MG onset. 
This introduces statistical artifacts.\footnote{To see this, as a null hypothesis, suppose that all jumps in social complexity occur during the interval from $-1.0$ to $1.0$ in PC1. Suppose as well that MG onset PC1 values
occur {independently}, by sampling a Gaussian centered on PC1 $= 0$. Finally, suppose that due to archaeological artifacts, half of all time-series have a continuous sequence of PC1 values stretching back only to PC1 $=1.0$, and the other half stretch back to 
before PC1 $= -1.0$. Then it will appear that there is probability $2/3$ of MG onset occurring after the jump in social complexity, even though the real probability is $1/2$.}

\begin{figure}
\centering
\includegraphics[width=.9\linewidth]{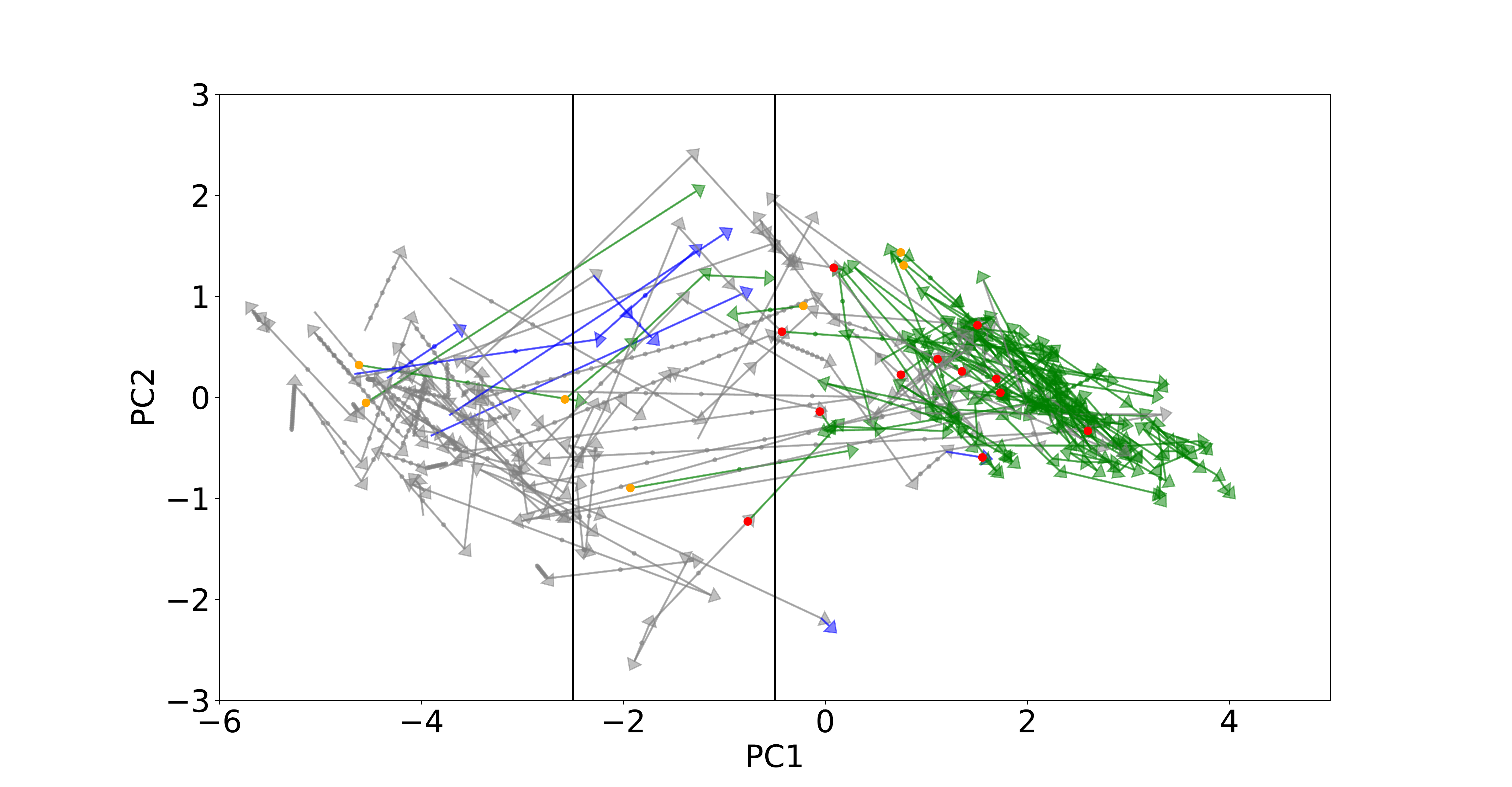}
\caption{TBD}
\label{fig:mg}
\end{figure}

The discovery of hinge points provides an alternative way to investigate the relationship between jumps in social complexity and the appearance of MGs, without these two shortcomings.
%
As shown in Figure~\ref{fig:mg}, MGs arise in polities both before or after the polity undergoes a ``jump in complexity''. At least based on the complexity thresholds found by analyzing Seshat
\citep{Shin2020}, there are instances in which MGs arise before the threshold at PC1 $= -2.5$, and instances in which MGs arise after the threshold at PC1 $= -0.5$. 

From a social science perspective, this suggests that a 
discontinuity in social complexity neither causes the onset of MGs nor requires them, contradicting several other analyses in the literature (including one that was based on the same Seshat dataset). In terms of building an underlying stochastic process model, these results are broadly consistent with the hypothesis that the onset of MGs is a jump of the first kind, where a large change of $\mathbf{x}(t)$ occurs with small but non-negligible probability under the transition matrix  $W_\theta(\mathbf{x} | \mathbf{x}')$.\footnote{Intuitively, this hypothesis is similar to supposing that the onset of MGs is a Poisson process, with a rate that is non-negligible across a broad span of PC1 values. We emphasize though that we have not yet done a proper statistical test of this hypothesis.} 

Note though that precisely because such a Poisson process is independent of the current value of $\mathbf{x}(t)$, as well as previous values, it is very challenging to distinguish the hypothesis that the onset of MGs is a Poisson process from the hypothesis that it is be an exogenous perturbation. This is a significant difference from the hinge points themselves. Because the thresholds captured in those hinge points \textit{do} depend on the current value of $\mathbf{x}(t)$, (by definition), and since the precise value of $t$ differs widely from one polity to another, those hinge points seem much more certain to be a jump, albeit of the second kind.

\section{A 1-dimensional random walk as a simple stochastic process}
We illustrate stochastic processes using perhaps the simplest such process there is. A random variable $x$ that at every discrete time step assumes some integer value, and from one time step to the next can change its value by $\pm 1$, The probability of going up is $p$, and that of going down hence $1-p$. Such jumps at different times are independent of each other. Only $x$, but not $p$ can be directly observed. $p$ can only be estimated from the observed data of $x$.  Suppose we observe a finite time series of points generated by sampling such a process, but cannot directly observe $p$. If $p$ changes at some time $t$ -- an exogenous perturbation that we do not observe -- then the dynamics after $t$ will be different. But based only on the time-series, no matter how different it may look before and after $t$, we cannot definitely conclude  that $p$ changed, either at $t$ or some other time. We can only suspect such a change when the time series starts to look qualitatively different. 
The best we can do is statistically estimate that such a change occurred. Fortunately, there do exist powerful techniques for such estimates.

Next, suppose that $p$ never changes, but we coarse-grain $x$, into bins of width $5$. Then if we know that the system is currently in the bin $\{0,1,2,3,4\}$, in order to predict the probability that at the next time step the system will be in bin $\{5, 6, 7, 8, 9\}$ we need to estimate the relative probabilities of which precise point in $\{0,1,2,3,4\}$ the system is currently in. In general, we will assign a non-zero probability to the event that the system is currently at the precise point $4$, and therefore assign a nonzero probability that at the next time step the system is in $\{5, 6, 7, 8, 9\}$. On the other hand, if we also know that at the previous time step the system was in the bin $\{-5,-4,-3,-2,-1\}$, then in fact we know the system is currently at the point $1$, with probability $1$, and so cannot be in the bin $\{5, 6, 7, 8, 9\}$ in the next time step. So knowing something about the past state of the system, in addition to knowing its current state, changes the relative probabilities of its future states. This illustrates that coarse-graining the observed time-series will in general change a Markovian dynamics into a non-Markovian one.

\section{Seshat: Global History Databank}

Begun in 2009, the Seshat project has pursued the ambitious goal of developing  a dataset to test theories about sociocultural evolution by cataloging the development of human civilization from the dawn of the Neolithic to the Industrial Revolution \cite{SeshatDB}. Two foundational elements of Seshat are the Natural Geographic Area (NGA) and polity. An NGA is a roughly 100 km by 100 km geographic area delimited by natural geographical features such as river basins, coastal plains, valleys, islands, and so forth \cite{turchin_etal2018}. A polity is an independent political unit that controls territory, and can range in size from small groups organized in local, independent local communities to territorially expansive, multi-ethnic empires. At any given time, exactly one polity controls an NGA, though that controlling polity may have its base or capital outside the NGA. For example, the Konya Plain NGA is located in the Central Anatolia Region of contemporary Turkey and has an area of 28,900 square km. It was controlled by the Hittite Empire (a polity) in 1000 BC and by the Eastern Roman Empire (also a polity) in AD 500. Seshat data used in recent analyses (i.e., the CCs in the World Sample 30; see below) are coded at 100-year  intervals. Thus, a useful way  to  think about the Seshat data is as a data matrix in which each row consists of an NGA-polity-time triplet. 

The Seshat database contains an ever-growing set of variables, now well over 1500, and cases,  coded through time for each polity in consultation with archaeological and historical experts. For example, for the Hittite Empire in 1000 BC, the following variables related to money have these codes:

\begin{center}
\textbf{Neo-Hittite Empire in 1000 BC (Population 1.3 - 2.0 million)}

\vspace{5mm} 
\begin{tabular}{ |c|c| } 

Articles & Present \\ 
Tokens & Absent  \\ 
Precious Metals & Present  \\
Foreign coins & ??  \\ 
Indigenous coins & ?? \\
Paper currency & Absent \\ 
$\cdots$ & $\cdots$ \\ 

\end{tabular}
\end{center}

For comparison, when the Konya Plain was controlled by the Eastern Roman Empire in AD 500, it had a population of about 15 million people. 

Seshat is constantly evolving as new data are added and old data re-assessed. This includes the addition of new variables and new NGAs. Recent articles that utilize the Seshat database have used a fixed version with 30 NGAs called the World Sample 30. NGAs for this sample were chosen to maximize geographic extent and diversity in social organization. In particular, 3 NGAs were chosen from each of 10 world regions (Africa, Europe, Central Eurasia, Southwest Asia, South Asia, Southeast Asia, East Asia, North America, South America, and Oceania-Australia) and the 3 NGAs in each region were selected such that sociopolitical complexity arrived relatively early, intermediate, and late. For this World Sample 30, recent analyses have also utilized a summary version of the dataset in which 51 variables were aggregated into nine distinct variables called ?Complexity Characteristics? (CCs):

\begin{enumerate}
  \item Population: Population of the entire polity
  \item Territorial Area: Territorial extent (area) of the polity
  \item Capital Population: Population size of the largest urban center (usually the capital)
  \item Hierarchical Levels: Number of types of settlements (e.g., hamlets to cities) and levels of administrative hierarchy
  \item Government: Aspects of government and bureaucracy, such as the presence of a legal code and merit promotion
  \item Infrastructure: Presence of bridges, roads, irrigation, etc.
  \item Information Technology/Writing: Presence and type of writing and recording systems
  \item Texts: Presence of specialized literature, including scientific texts, histories, calendars, fiction, etc.
  \item Money: The monetary system---presence of local/foreign currencies, paper currency, tokens of exchange, etc. (see Neo-Hittite example above)

\end{enumerate}

Each of the CCs is normalized to lie between 0 and 1, and for each polity-time pair in the World Sample 30 there exists an observation for each CC. One challenge of working with the Seshat data is that, especially for earlier polities, there is insufficient evidence to code many variables---and often disagreements among experts. 

Rather than assign a value for each CC and create one canonical imputation, the Seshat team assigned a distribution of possible values. These distributions are then sampled 20 times, to produce 20 replicates of the dataset \cite[p. 7]{turchin_etal2018}. The imputation is performed on each replicate, producing 20 different sets of CC values. These replicates are used to produce confidence intervals on the proportion of variance explained by each PC, the component loadings, and the values of the PCs for each polity. 

\section{A hidden Markov model of age-specific demography}

Consider an age structured population of reproductive females (including juveniles) where the age-class $j$ contains individual between $\Delta a \cdot j$ and $\Delta a \cdot (j+1)$ years of age, with $\Delta a$ being the age-spacing; it is straightforward to generalize this to males and post-reproductive females, as well as to account for other types of population structure, such as spatial location and socioeconomic status\cite{rogers1966,caswell2001}. The column vector $\mathbf{z}_t$ gives the number, or proportion, of females in each age-class at time step $t$. $F_j$ is the age-specific fertility of females in age-class $j$, accounting only for female offspring. The age-specific survival of females in age-class $j$ is $P_j$. Given these definitions, the population project matrix for reproductive females is
\begin{equation}
	\label{eq:A}
	\mathbf{A} = 
	\left( \begin{array}{cccccc}
		F_0 & F_1 & F_2 & F_3    & \cdots  & F_{J}   \\
		P_0 & 0   & 0   & 0      & \cdots  & 0       \\
		0   & P_1 & 0   & 0      & \cdots  & 0       \\
		0   & 0   & P_2 & 0      & \cdots  & 0       \\
		0   & 0   & 0   & \ddots & \cdots  & 0       \\
		0   & 0   & 0   & \ddots & P_{J-1} & 0
	\end{array} \right)
	\mbox{,}
\end{equation}
where $J$ is the last reproductive age class. The population project equation is
\begin{equation}
	\label{eq:pop_proj}
	\mathbf{z}_{t+1} = \mathbf{A} \, \mathbf{z}_t \mbox{.}
\end{equation}
\noindent These preceding equations succinctly summarize three effects: Females get older, that is, advance from one age class to the next, give birth to young females, and may die.

\textbf{Stable demography}: If demographic rates are stable\footnote{Certain technical constraints on $\mathbf{A}$ must also be satisfied. In particular, $\mathbf{A}$ must be irreducible and primitive, which is almost always true.} the population vector $\mathbf{z}_t$ converges to a constant growth rate and stable age distribution. The growth ``rate'' (more precisely, the per period growth factor) equals $\lambda$, the dominant left eigenvalue of $\mathbf{A}$. The stable age distribution ($\mathbf{u}$) equals the corresponding dominant right eigenvector and the age specific reproductive value ($\mathbf{v}$) equals the corresponding left eigenvector \cite{fisher1958,caswell2001}.\\

\noindent \textbf{Time dependence}: Rather than assuming a constant population projection matrix, we now add a subscript to allow time-dependence, $\mathbf{A}_t$. The time-dependent population projection equation is 
\begin{equation}
	\label{eq:pop_proj_time}
	\mathbf{z}_{t+1} = \mathbf{A}_t \, \mathbf{z}_t \mbox{.}
\end{equation}



\noindent One can easily expand the preceding model to a full demographic model that includes males and post-reproductive females. Parameters such as the sex-ratio at birth and age-specific mortality for the additional segments of the population can be inferred from known demographic statistics and/or observed data. 
The full demographic specification for time period $t$ is $\mathcal{D}_t = \{ \mathbf{F}_t,\mathbf{P}_t\}$, where $\mathbf{F}_t$ is the time-dependent vector of age-specific fertilities and $\mathbf{P}_t$ is the time-dependent vector of age-specific survivals, and the demographic model for the period $0$ to $t$ is $\mathcal{G}_t = \{\mathcal{D}_0, \mathcal{D}_1, \cdots, \mathcal{D}_{t-1}\}$. An implicit assumption in this definition is that the population vector at time step $0$ is set assuming stable demography defined by the demographic specification $\mathcal{D}_0$. This assumption could be relaxed by letting $\mathbf{z}_0$ be a free parameter in the demographic model $\mathcal{G}_t$.

\subsection{From demographic model to likelihood and a hidden Markov model}
To link the preceding model to the stochastic process framework suggested in the main text, we now interpret the population projection matrices $\mathbf{A}_t$ as being determined by latent variables in a hidden Markov model. Further, we assume that there exists an external variable, let's say for the sake of concreteness a categorical climate variable (likely slowly changing) that can be directly or indirectly observed, indexed by $k=1 \cdots K$, where for each climate state, $k$, different transition probabilities apply:

\begin{equation}
	\label{eq:transitions_k}
	\mathbf{p}_t = \left[ \sum_{k=1}^{K} c_k \, \mathbf{W}^{(k)} \right] \mathbf{p}_{t+1} \mbox{,}
\end{equation}

\noindent where $c_k$ is an indicator variable for the categorical climate state and $\mathbf{W}^{(k)}$ is the transition matrix that applies for climate state $k$. There are many alternatives to this approach, including modeling the projection matrices $\mathbf{A}_t$ as depending on a continuous, exogenous climate parameter vector $\mathbf{\theta}$. However, the model we describe is sufficiently simple to be motivating yet sufficiently complex to be realistic. We assume that there exists a set of $N$ reference dynamics indexed by $n = 1 \cdots N$, where each reference dynamics, $\mathcal{D}^{(n)}$, is for a distinct annual demographic state -- for example, one $n$ could correspond to famine, another warfare, etc. Next define the vector

\begin{equation}
	\label{eq:q}
	\mathbf{q} = 
	\left[ \begin{matrix} m_1 & m_2 & \cdots & m_T \end{matrix} \right]^T
\end{equation}

\noindent to be the population dynamics state, $n$, for the time periods $1$ through $T$. Given the preceding formulation, it is straightforward to calculate the probability of any given $\mathbf{q}$ given the set of transition probabilities (to be inferred) and the set of reference dynamics (pre-specified); we denote this probability by $p(q|\{\mathbf{W}^{(k)}\},\{\mathbf{A}^{(n)}\})$, where we indicate the set of transition matrices with $\{\mathbf{W}^{(k)}\}$ and the set of reference population project matrices with $\{\mathbf{A}^{(n)}\}$. One can then sum over the set of valid vectors $\mathbf{q}$ and use Equations~\ref{eq:A} and \ref{eq:pop_proj} to calculate the probability of each $\mathbf{z}_t[j]$. Finally, this can be linked to available archaeological data to calculate a likelihood function, which can be used for either maximum likelihood estimation or Bayesian inference. For example, given a set of radiocarbon determinations  (e.g., as in \cite{price_etal2020}) one can assume that the probability a given sample is from a given year is proportional to the total population size in that year, the sum of the elements of $\mathbf{z}$. Similarly, if skeletal age-at-death is known, and a rough date estimate is available from associated artifacts, one can calculate the relative probability of being in the pertinent age-class (and sex-class if the model is suitably generalized) from the population vectors by summing across years as determined by the associated artifacts. Naturally, one can use multiple types of data as part of a single likelihood calculation to improve inference of the underlying model parameters; a major benefit of the approach we have described is that it is straightforward to accommodate additional types of data in the likelihood calculation in order to further improve inference (e.g., isotopic data to inform on migration, health data such as from linear enamel hypoplasias (LEHs) to improve inference on mortality, and both ancient and modern genetic data).

\bibliography{./main.tweaked.to.compile,./supp.tweaked.to.compile}

\end{document}